\documentstyle[12pt,aaspp4,epsf]{article}
\righthead{Sodium and Aluminum in Red Giants}
\begin{document}

\title{The Production of Sodium and Aluminum in Globular Cluster Red
Giant Stars}
\author{Robert M. Cavallo}
\affil{Department of Astronomy, University of Maryland, College Park, MD
20742; rob@astro.umd.edu}
\authoraddr{Department of Astronomy, University of Maryland, College Park, MD
20742; rob@astro.umd.edu}
\author{Allen V. Sweigart}
\affil{Laboratory for Astronomy and Solar Physics, Code 681, NASA/Goddard
Space Flight Center, Greenbelt, MD, 20771; sweigart@bach.gsfc.nasa.gov}
\authoraddr{Laboratory for Astronomy and Solar Physics, Code 681, NASA/Goddard
Space Flight Center, Greenbelt, MD, 20771; sweigart@bach.gsfc.nasa.gov}
\and
\author{Roger A. Bell}
\affil{Department of Astronomy, University of Maryland, College Park, MD
20742; roger@astro.umd.edu}
\authoraddr{Department of Astronomy, University of Maryland, College Park, MD
20742; roger@astro.umd.edu}

\begin{abstract}
	We study the production of Na and Al around the hydrogen shell of two
red-giant sequences of different metallicity in order to explain the abundance
variations seen in globular cluster stars in a mixing scenario.
Using detailed stellar models together with an extensive nuclear reaction
network, we have calculated the distribution of the various isotopic abundances
around the hydrogen shell at numerous points along the red-giant branch.
These calculations allow for the variation in both temperature and density
in the shell region  as well as the timescale of the nuclear processing, as
governed by the outward movement of the hydrogen shell.
The reaction network uses updated rates over those of Caughlin \& Fowler
(1988).
We find evidence for the production of Na and Al occurring
in the NeNa and MgAl cycles.  In particular, Na is significantly enhanced
throughout the region above the hydrogen shell.
The use of the newer reaction rates causes a substantial increase in
the production of $^{27}$Al above the hydrogen shell through heavy
leakage from the NeNa cycle and should have an important effect on the
predicted surface abundances.  We also find that the nuclear processing is
considerably more extensive at lower metallicities.

\end{abstract}


\section{Introduction}

	Briley et al. \markcite{r5}(1994) and Kraft \markcite{r6}(1994)
have reviewed the observational data on variations in the abundances of
C, N, O, Na, and Al in globular cluster red-giant-branch (RGB) stars.
The variations in Na and Al are particularly significant, since they have
long been regarded as one of the principal arguments in favor of a primordial
origin for the abundance anomalies (see e.g., Cottrell \& Da Costa
\markcite{r19}1981).
Star-to-star variations of Na were first observed by Cohen \markcite{r1} (1978)
and Peterson \markcite{r2}(1980) in M13 and M3, while similar variations of Al,
which are correlated with the CN band strength,
were found by Norris et al. \markcite{r3}(1981) in NGC 6752.
Since these original observations, numerous other groups have confirmed the
general existence of Na and Al vs. N correlations and Na and Al vs. O
anticorrelations in globular cluster red giants
(Drake et al. \markcite{r24}1992;
Kraft et al. \markcite{r25}1992, \markcite{r26}1993).
Recently Norris \& Da Costa \markcite{r7}(1995) have concluded
that Na variations exist in all clusters,
while Al variations are greater in the more metal-poor clusters.

	Except for the modest alterations due to the first dredge-up,
canonical stellar evolution models predict no changes in the C, N, O, Na,
and Al surface abundances during the RGB phase.
In an attempt to explain this discrepancy between the predicted and observed
abundance variations, Sweigart \& Mengel
\markcite{r8}(1979; hereafter, SM79) suggested that meridional circulation
currents, driven by internal rotation, might be able to mix material across
the radiative zone that separates the top of the hydrogen shell (H shell) from
the base of the convective envelope in canonical RGB models.
They found that there was a region of significant extent just above the H shell
within which the CN cycle has processed C into N and, somewhat closer to the
shell, there was a region within which the ON cycle has processed O into N.
The mixing was postulated to begin at the point along the RGB where the H shell
burns through the hydrogen discontinuity that was previously produced by the
deep penetration of the convective envelope during the first dredge-up.
A progressive depletion of carbon is seen along
the giant branch in M92, M15, and NGC 6397 (Bell, Dickens, \& Gustafsson
\markcite{r23}1979; Carbon et al. \markcite{r27}1982; Trefzger et al.
\markcite{r28}1983; Briley et al. \markcite{r29}1990), although the luminosity
at which this depletion begins is uncertain, for observational reasons.
If Na and Al were also manufactured in the CN- and ON-processed
regions, then mixing might also explain the observed Na and Al variations.
Complimenting the SM79 hypothesis, Denisenkov \& Denisenkova \markcite{r9}
(1990; hereafter, DD90) have suggested $^{23}$Na can be produced from
proton captures on $^{22}$Ne in the ON-processed region and
that the rotation rate required by SM79 is sufficient to
reconcile the observations with theory.

      Expanding on the concepts of SM79 and DD90, Langer, Hoffman, \& Sneden
\markcite{r10}(1993; hereafter, LHS93; see also Langer \& Hoffman
\markcite{r31} 1995 and Denissenkov \& Weiss \markcite{r32} (1996))
examine the production of Na and Al in RGB stars by following the reactions
of the relevant nuclei in a simplified model with a low metallicity
(Z = 0.0001 or [Fe/H] = -2.3) and a constant temperature (T$_{9}$ = 0.040 where
T$_{9}$ = T/10$^{9}$K) and density ($\rho$ = 44.7 g cm$^{-3}$).
Their study shows a significant enhancement of $^{23}$Na in the ON-processed
region which derives from a series of proton captures on
$^{20}$Ne in the NeNa cycle.
Furthermore, their model yields an increase in $^{27}$Al which is made
from proton captures on both $^{25}$Mg and $^{26}$Mg via the reactions
$^{25}$Mg$(p,\gamma)^{26}$Al$(\beta^{+})^{26}$Mg$(p,\gamma)^{27}$Al.
Their study also shows that a MgAl cycle is set up after 1.7 Myr.
This cycle begins with a proton capture on $^{24}$Mg, which is itself
enhanced through leakage from the NeNa cycle by a $(p,\gamma)$ reaction
with $^{23}$Na, and is completed with $^{27}$Al$(p,\alpha)^{24}$Mg.
Both the NeNa and MgAl cycles are depicted graphically in Figure 1.

\begin{figure}
\plotone{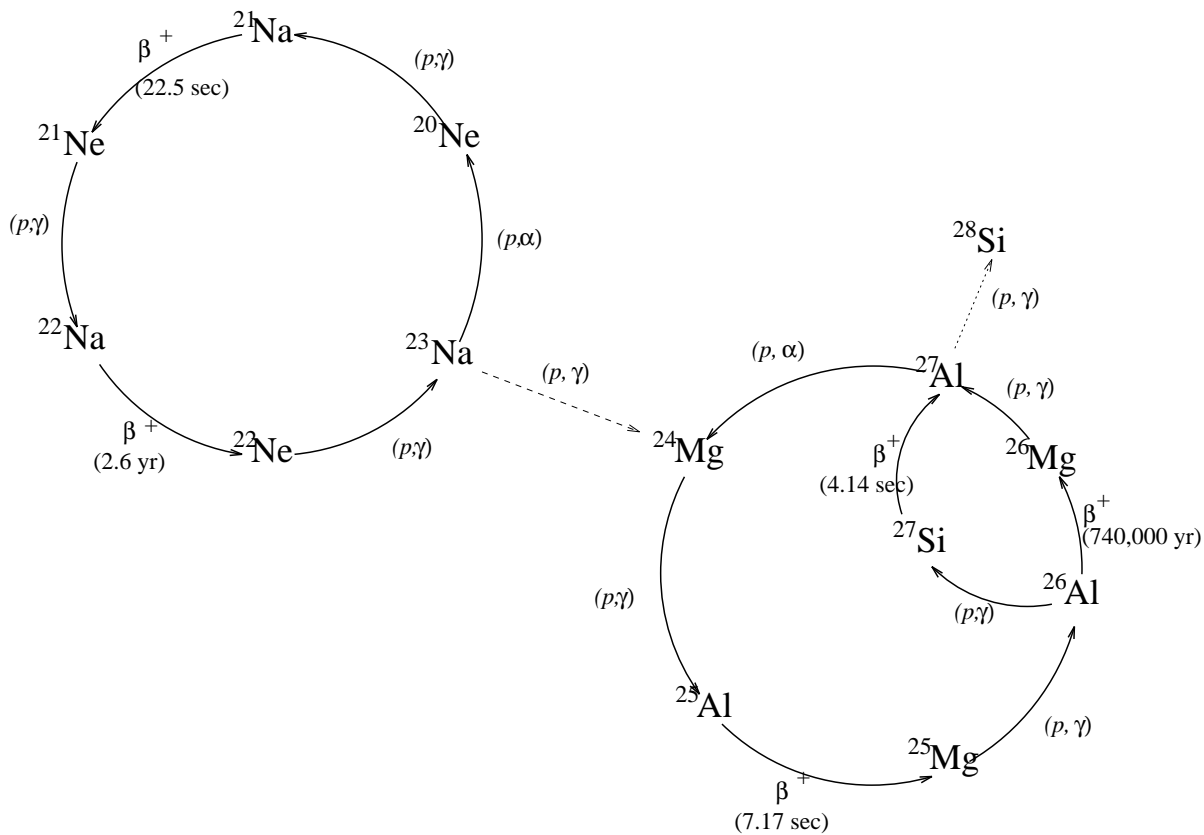}
\caption{The reactions involved in the NeNa and MgAl cycles.
The half-lives and reaction types are given parenthetically.
In the case of competing decay paths, the solid lines show the stronger path,
derived with the rates used in our code for the relevant temperature range.
The dashed lines show the weaker decays which lead to a depletion in the
total abundance of each cycle ("leakage").}
\end{figure}

The purpose of this ${\it Letter}$ is to develop the work
of LHS93 further by using detailed stellar evolutionary sequences to examine
the production of Na and Al around the H shell in RGB stars.
This represents an improvement over LHS93 because: 1) we incorporate the
latest reaction rates in our burning code, 2) we explore a wider parameter
space in both metallicity and luminosity, and 3) our more realistic RGB models
take into account the variation in temperature and density around the H shell
and incorporate the timescale for the nuclear processing to produce Na
and Al, as set
by the rate at which the H shell moves outward in mass.  The Na and Al
produced in this fashion can then be mixed outward into the envelope over
the relevant timescale for the mixing process.
Section 2 summarizes our numerical techniques and input physics while
section 3 presents the results of our calculations for some representative
cases.  We conclude with a brief discussion of our results in section 4.

\section{Sequences and Network}

      In this ${\it Letter}$ we present results for two RGB sequences,
one with mass M = 0.795 M$_\odot$ and scaled-solar metallicity Z = 0.0001,
and one with M = 0.875 M$_\odot$  and Z = 0.004.   Each of these sequences
was evolved from the zero-age main sequence up the giant branch to the onset
of the helium flash.  The masses were chosen to give an age of 15 Gyr
at the helium flash.  We began our calculation of the abundance
distributions around the H shell at the point on the RGB where the H shell
burns through the hydrogen discontinuity produced during the first dredge-up.
This occurs at a luminosity of
log($\it {L/L}_\odot$) = 2.28 and 1.72 in the low and high metallicity
sequences, respectively.  Above this luminosity, mixing into the nuclearly
processed region at the top of the H shell is no longer hindered by a gradient
in the mean molecular weight.  Calculations of the abundance distributions
were carried out in increments of 0.005 M$_\odot$ in the helium-core mass
until the models reached the helium flash.
Typical temperatures and densities within the nuclearly processed region were
in the range from
0.015 $<$ T$_9 <$ 0.06 and 0.3 $<$ log$(\rho) <$ 2.5.

      Our nuclear reaction network code was kindly supplied to us by Dr. David
Arnett. The majority of the reaction rates are taken from Caughlin \& Fowler
\markcite{r11}(1988; hereafter, CF88).
When available, we include the more recent rates for the NeNa
and MgAl cycles as reviewed by Arnould, Mowlavi, \&
Champagne \markcite{r14}(1995; hereafter, AMC95).
In particular, we include the latest results for the Ne and Na proton
capture rates by El Eid \& Champagne \markcite{r15}(1995).
The Mg and Al proton capture rates are obtained from Iliadis et al.
\markcite{r16}(1990) and Champagne, Brown, \& Sherr \markcite{r17}(1993).
AMC95 re-evaluate the $^{27}$Al proton capture decay paths and
find, contrary to the results of Timmermann et al. \markcite{r18}(1988) and
Champagne et al. \markcite{r20}(1988) but in agreement with CF88,
that the $\alpha$-decay channel is stronger than the $\gamma$-decay channel,
so that the MgAl reaction rates do, in fact, lead to cycling.

	 The network code requires a run of temperature and density,
which are given by the models, and a timestep, which we supply
using the stationary shell approximation (see e.g. Sweigart
\markcite{r12}1994).
This approximation shows excellent agreement with the conventional solution
of the differential equations for the CNO reactions (Clayton
\markcite{r22} 1983)
and will be discussed more fully in a future publication.
The initial abundances for the main CNO isotopes outside the H shell
are obtained from the RGB models.
We input the initial values for the other elements from $^{19}$F to
$^{44}$Ca according to their scaled solar values (Anders \& Grevesse
\markcite{r13}1989).

\section{Results and Discussion}
\subsection{Results Using Updated Reaction Rates}

      Figure 2 shows the elemental profiles around the H shell of the low
metallicity sequence at the start of mixing and at the onset of the helium
flash.
Panels  a and b clearly demonstrate the initial rise in $^{23}$Na from
$^{22}$Ne throughout a region above the H shell, as predicted by DD90,
followed by the cycling of $^{20}$Ne into $^{23}$Na somewhat closer
to the shell, as discussed by LHS93.  This rise in $^{23}$Na occurs
within the region of C depletion which extends out to $\Delta \rm{M_r}$
= 0.0040 and 0.0015 in panels a and b, respectively.  The lower panels
highlight the MgAl cycle. As one progresses inward toward the H shell
in panel c, $^{26}$Al is produced first from $^{25}$Mg and
then converted into $^{27}$Al through the paths shown in Figure 1.
With advancement up the RGB, some $^{27}$Al is eventually destroyed via
$^{27}$Al$(p,\alpha)^{24}$Mg.
By the tip of the RGB, the
NeNa cycle (panel b) is leaking heavily into the MgAl cycle and producing
sizable enhancements in $^{27}$Al just above the H shell (panel d).
This huge increase in $^{27}$Al happens during the last 15\% of
the RGB lifetime,  possibly allowing enough time for mixing to the surface.

\begin{figure}
\plotone{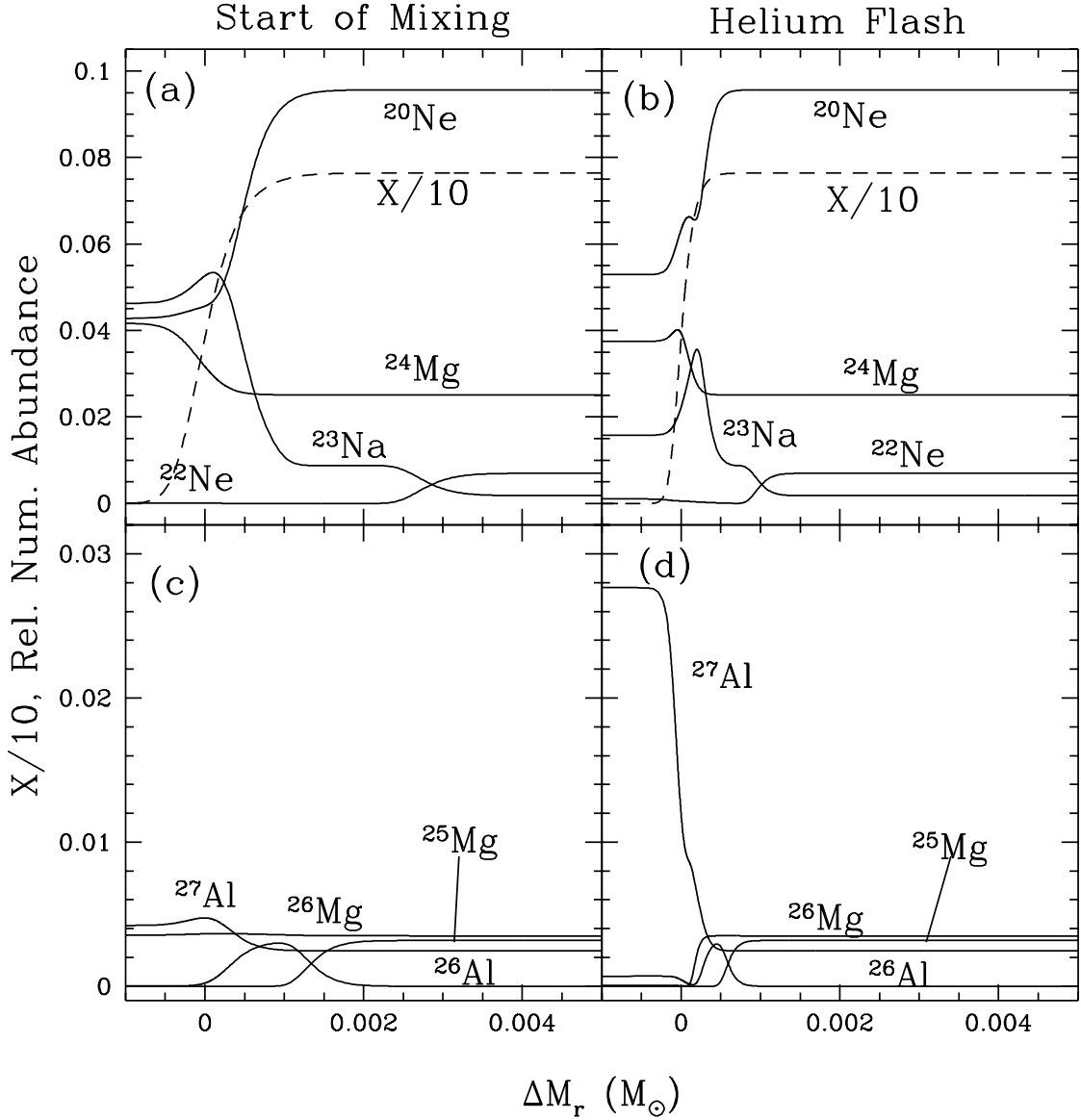}
\caption{Abundance profiles around the H shell for the
sequence with M = 0.795 M$_\odot$ and Z = 0.0001 at the start of mixing (left
panels) and at the onset of the helium flash (right panels).
The abscissa is the difference in mass from the center
of the H shell while the ordinate is the number abundance relative
to all metals. The hydrogen mass fraction, scaled to 0.1 of its actual
value, is given by the dashed line in the upper two panels.
The region of hydrogen depletion defines the H shell.}
\end{figure}

      For comparison, Figure 3 shows the same abundance profiles for the M =
0.875 M$_\odot$, Z = 0.004 sequence.
Since the temperature around the H shell is lower in this sequence,
the changes in the
abundances are less dramatic throughout the evolution up the RGB, when
compared to the low metallicity sequence.
Nevertheless, $^{23}$Na is still enhanced in a region above the shell,
especially at the start of mixing (panel a). Significant production of
$^{27}$Al, however, is only seen in panel d and even then only within the
H shell.  A similar decrease in the extent of the CN- and ON-processed regions
in RGB models of higher metallicities was found by SM79 .  Thus mixing in
more metal-rich globular cluster giants should have less effect on the
surface abundances, especially in the case of Al.  Some of these rough
estimates seem to be confirmed by Norris \& Da Costa \markcite{r7} (1995),
who note that the $\omega$ Cen giants with low [Fe/H] show a
greater production of Al relative to Na.

\begin{figure}
\plotone{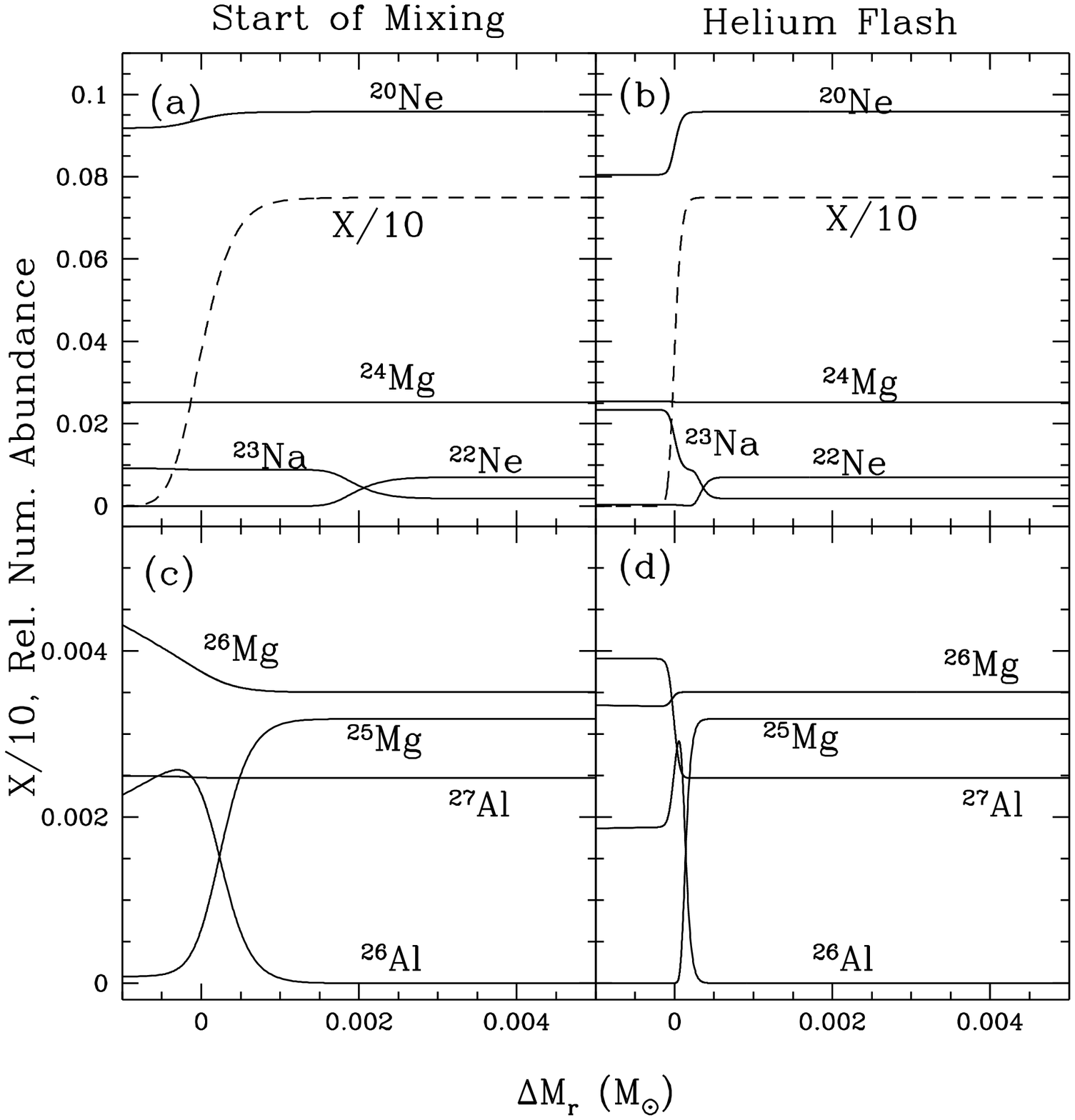}
\caption{As Figure 2, with M = 0.875 M$_\odot$ and
Z = 0.004.}
\end{figure}

\subsection{Comparison with the Caughlin \& Fowler (1988) Reaction Rates}

	In order to study the effects of using updated
reaction rates, we recalculated our profiles using the older CF88 rates.
These are plotted in Figure 4 for the low metallicity sequence.
The patterns indicative of NeNa and MgAl cycling are clearly visible.  The most
notable difference from the previous results is in the enhancement of $^{27}$Al
at the onset of the helium flash.  With the CF88 rates (Figure 4d), the
increase of $^{27}$Al at the center of the H shell is a factor
of two lower than with the newer rates (Figure 2d).  Further,
the large increases in Al only occur during the last 5\% of the RGB
lifetime.  With the CF88 rates, the NeNa cycle does not leak as heavily into
the MgAl cycle, resulting in a larger build-up of $^{23}$Na (Figure 4b
vs. Figure 2b) and a smaller enhancement of $^{24}$Mg.  This effect is offset,
however, by the fact that the region of $^{23}$Na enhancement extends
further above the H shell with the new rates.
Since mixing is postulated to occur only above the H shell,
one expect the more modern rates to increase the predicted surface abundances
of Na and Al.

\begin{figure}
\plotone{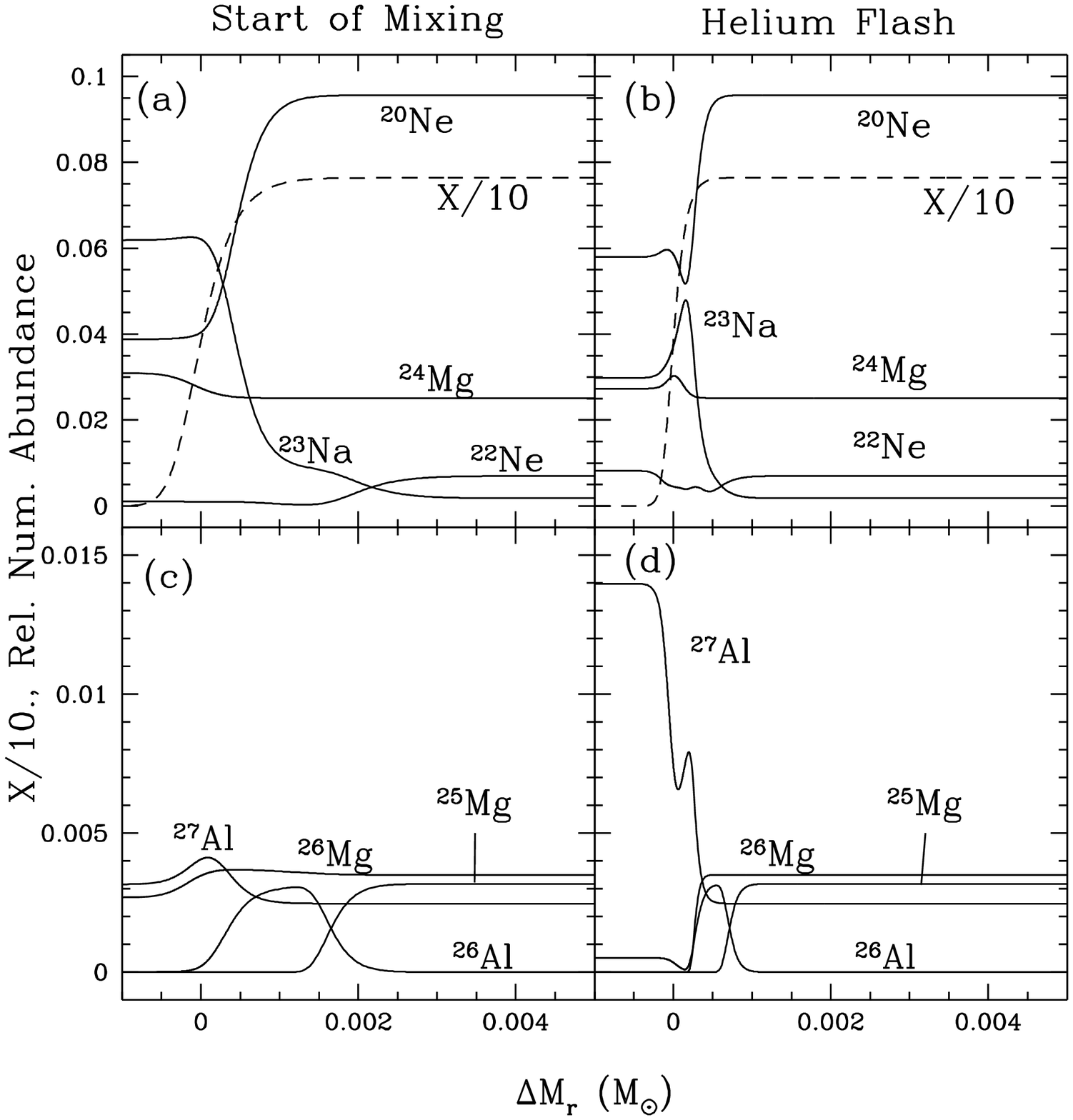}
\caption{As Figure 2, with the older rates of CF88. Note
the difference in the vertical scale in panels c and d when compared to
figure 2.}
\end{figure}

\section{Conclusion}

	By combining our realistic stellar models with an updated nuclear
reaction network, we are able to follow the nucleosynthesis of
C, N, O, Ne, Na, Mg, and Al around the H shell of two sequences of
differing metallicity as they evolve up the RGB.
In qualitative agreement with the observations, our results show an increase in
$^{23}$Na above the H shell throughout the entire giant branch, independent
of metallicity.  Furthermore, we produce sizable enhancements of $^{27}$Al
for the low metallicity sequence as it approaches the tip of the RGB,
without having to increase the initial $^{25,26}$Mg abundance or
the $^{26}$Mg proton
capture rate as in Langer \& Hoffman \markcite{r31} (1995).
Thus, our results can potentially reconcile the Na and Al abundance anomalies
observed in globular cluster giants with the mixing of these elements from
the stellar interior.

	Although our work is based on up-to-date reaction rates,
those rates can still be quite uncertain in some cases.
For example, our results deviate from the observations in the case of
$^{24}$Mg.
Contrary to the findings of Shetrone \markcite{r30}(1996), who observes
a decrease in $^{24}$Mg with increasing luminosity in M13 giants, we show
an enhancement of $^{24}$Mg with continuing evolution for the low metallicity
sequence.  However, the high metallicity sequence shows only a marginal
increase  in the $^{24}$Mg abundance.
Thus, our present inability to qualitatively reproduce the
$^{24}$Mg observations might be due to the current uncertainty
in the nuclear cross-sections.
Perhaps a faster rate for the $^{24}$Mg$(p,\gamma)^{25}$Al reaction or a
slower rate for the $^{27}$Al$(p,\alpha)^{24}$Mg reaction would
decrease the $^{24}$Mg abundance.  Figures 2 and 4 show how the $^{24}$Mg
abundance can depend on the choice of reaction rates.

	In addition to depending on the accuracy of the nuclear reaction rates,
any final predictions of surface abundances will also depend strongly
on the choice of a mixing mechanism.  The mixing timescale, the depth of
the mixing, and the dependence on metallicity will all affect the
quantitative results for the surface abundances.  In a future paper, we will
explore these aspects of the mixing mechanism and will compare the predicted
variations of the surface abundances along the RGB with the observational
constraints.

	In order to better understand the effects of metallicity,
we will also extend the present work to a  Population I red giant.
Furthermore, we will use the same set of sequences to study the CNO isotopes.
Finally, we intend to
follow the production and destruction of $^{3}$He along the RGB in order
to determine better how low mass stars affect the chemical evolution of
this isotope in the Galaxy.

\acknowledgments
	The authors wish to thank Dr. David Arnett of the University of
Arizona for the use of his fine nuclear reaction network and for his
immeasurable patience in helping us implement the code.
We also wish to thank Drs. M. Arnould, N. Mowlavi, and A.
Champagne for allowing us to preview their work on the uncertainties in
our reaction rates.
The work of A.V.S. is funded in part by NASA RTOP 188-41-51-03 and that of
R.A.B. by NSF AST93-14931.
Finally, R.M.C. wishes to acknowledge the NASA Graduate Student
Research Program for financial support of his research.

\end{document}